# Noise Reduction Methods for Large-Scale Intensity Mapping Measurements with Infrared Detector Arrays

Grigory Heaton,[1] Walter Cook,[1] James Bock,[1] Jill Burnham,[1] Sam Condon,[1] Viktor Hristov,[1] Howard Hui,[1] Branislav Kecman,[1] Phillip Korngut,[1] Hiromasa Miyasaka,[1] Chi Nguyen,[1] Stephen Padin,[1] and Marco Viero[1]

[1]*California Institute of Technology, Cahill Center for Astronomy and Astrophysics, 1216 East California Boulevard, Pasadena, CA 91125*

## ABSTRACT

Intensity mapping observations measure galaxy clustering fluctuations from spectral-spatial maps, requiring stable noise properties on large angular scales. We have developed specialized readouts and analysis methods for achieving large-scale noise stability with Teledyne 2048×2048 H2RG infrared detector arrays. We designed and fabricated a room-temperature low-noise ASIC Video8 amplifier to sample each of the 32 detector outputs continuously in sample-up-the-ramp mode with interleaved measurements of a stable reference voltage that remove current offsets and $1/f$ noise from the amplifier. The amplifier addresses rows in an order different from their physical arrangement on the array, modulating temporal $1/f$ noise in the H2RG to high spatial frequencies. Finally, we remove constant signal offsets in each of the 32 channels using reference pixels. These methods will be employed in the upcoming SPHEREx orbital mission that will carry out intensity mapping observations in near-infrared spectral maps in deep fields located near the ecliptic poles. We also developed a noise model for the H2RG and Video8 to optimize the choice of parameters. Our analysis indicates that these methods hold residual $1/f$ noise near the level of SPHEREx photon noise on angular scales smaller than $\sim 30$ arcminutes.

## 1. INTRODUCTION

The history of galaxy formation can be studied using intensity mapping measurements of the near-infrared extragalactic background light. Galaxy clustering produces large-scale structure on the order of tens of arcminutes (Cooray et al. (2004), Kashlinsky et al. (2004)). The sources that comprise the background can be accessed using fluctuation measurements rather than galaxy-by-galaxy photometry (e.g. Zemcov et al. (2014)). Such intensity-mapping measurements promise to trace total light production, including diffuse and faint sources of emission, over the history of galaxy formation back to the Epoch of Reionization (Kovetz et al. (2017)). In order to optimize measurements of these large-scale background fluctuations, it is critical to minimize readout noise on tens of arcminute scales in astrophysical images.

The Spectro-Photometer for the History of the Universe, Epoch of Reionization, and Ices Explorer (SPHEREx) mission is an all-sky infrared (0.75-5.0 µm) survey satellite (Doré et al. (2014), Crill et al. (2020)) selected under NASA's Medium Explorer (MIDEX) program. Alongside other mission objectives, in deep maps, SPHEREx will study large-scale infrared extragalactic background structure. SPHEREx uses 6 Teledyne HAWAII-2RG (H2RG) HgCdTe sensors (Blank et al. (2012)), each with a 2048×2048 array of pixels. Included in this count is a frame of reference pixels which are not sensitive to light, consisting of the outermost 4 pixels surrounding the sensor. These reference pixels are designed to emulate the electrical response of a typical optical pixel, allowing them to be used to correct for dark current and electrical noise in an image. The array is divided into 32 2048×64 pixel channels, each of which is read out simultaneously row by row. Noise from the readout electronics can manifest itself as spatial $1/f$ noise when the H2RG array is read.

In order to minimize spatial noise on angular scale $\theta \approx 30'$ in intensity mapping observations, we present several new methods for reducing the effects of electronics $1/f$ noise. "Row-chopping" reduces the effect of temporal $1/f$

Corresponding author: Grigory Heaton
gheaton@caltech.edu

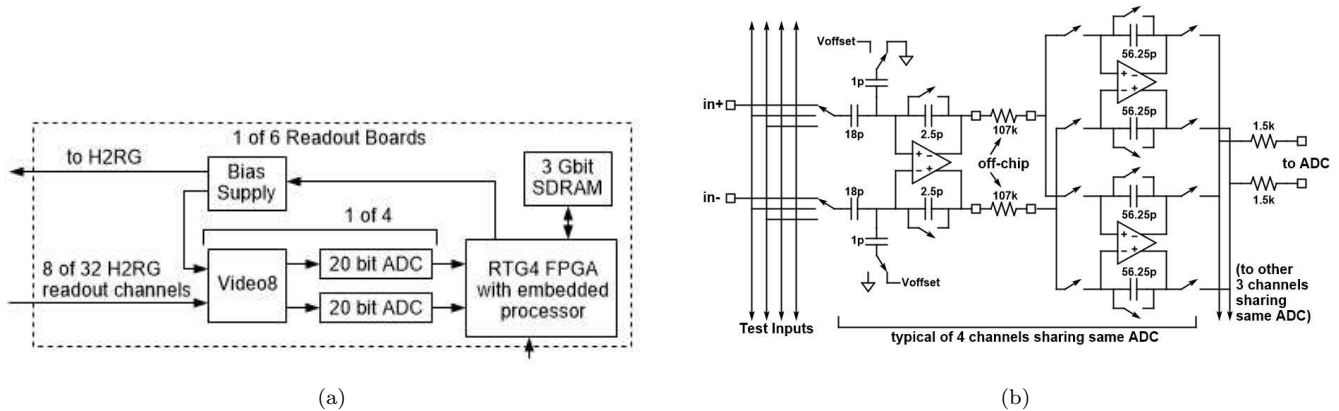

(a) (b)

**Figure 1.** (a) Overview of the SPHEREx readout electronics and (b) internal schematic of a single Video8 preamplifier channel, one of eight on the chip. Differential input lines are capacitively coupled to a low-noise integrator, and the output is read with one of two sample and hold circuits. The Video8 input can be switched to a reference voltage intermittently to track amplifier drift.

noise by reading rows non-sequentially. Combined with multiple H2RG reference pixel samplings, we obtain significant noise reduction on the spatial scales of interest, optimizing large-scale background fluctuation measurements. We use a phantom pixel correction scheme based on measuring a stable reference voltage to remove dark current offsets produced by our custom ASIC amplifier. We validated these noise reduction methods first in a noise simulator and later in laboratory tests with a physical H2RG array.

In this paper, spatial noise power is expressed in terms of the spatial power spectrum $P(k)$, which represents the azimuthally-averaged spatial power as a function of angular wave number. For SPHEREx detectors, each array pixel corresponds to 6.2 arcseconds of sky angle, resulting in a wave number $k = 2\pi(6.2''/\theta)$ in pix$^{-1}$ units for an angular period $\theta$. This paper is structured as follows: In Section 2 we describe the SPHEREx readout electronics and Video8 amplifier, and outline various noise reduction techniques. In Section 3 we describe a full noise simulator to model the electronics noise, followed by optimization of noise reduction techniques in Section 4. Section 5 compares the simulation outputs and modeled noise reduction with measured laboratory images. Finally, Section 6 summarizes our conclusions.

## 2. SPHEREX READOUT ELECTRONICS

### 2.1. Readout Boards

The SPHEREx readout electronics consist of 6 identical readout boards (see Figure 1), a central electronics board (CEB), and a low voltage power supply. The boards are housed in a card cage and communicate with each other and the spacecraft via a backplane. Each readout board services a single H2RG array, providing bias, control signals, and readout via low-noise Video8 amplifiers (see Section 2.2) operated in 32-channel 100kHz mode. The channel blocks are 64 pixels wide (see Figure 2). The readout board logic and processor are embedded in an RTG4 FPGA, with a 3 Gbit SDRAM device providing on-board sample-up-the-ramp processing for all $\sim 4$ million H2RG pixels. The processor, aided by dedicated logic, compresses and formats the sample up the ramp data before sending it to the spacecraft via the CEB. Each readout board also has a housekeeping system to monitor various voltages and temperatures.

The CEB interfaces with the spacecraft, communicates with the 6 readout boards, monitors 25 Cernox temperature sensors mounted on the telescope, and controls the temperature of the focal plane arrays via precision readout of 8 Cernox sensors and 16 bit control of 8 heaters. An RTG4 FPGA with an embedded processor orchestrates the CEB activity, which includes the generation of low-speed (38.4 kbaud) housekeeping telemetry and the booting of all the embedded processors using code stored in MRAM.

The H2RG bias supply is designed for stability and is housed in an oven-controlled region of the readout board to minimize thermal drifts. Each of the 32 H2RG channels sees a common bias voltage from this supply, resulting in any bias noise fluctuations or temperature offsets being repeated across all channels.

## 2.2. *Video8 Amplifier*

The Video8 amplifier is an Application-Specific Integrated Circuit (ASIC) designed at Caltech for SPHEREx. It interfaces preamplifiers and integrators between SPHEREx H2RG arrays and external RT2378-20 20 bit ADCs. The Video8 amplifiers are designed for stable operation over a wide temperature range from -50°C to +50°C. For SPHEREx, the Video8 devices are located within the instrument electronics box and do not require special cooling. A cryo-harness connects the H2RG detectors, operating at 50-80K, to the Video8 inputs. In order to minimize spatially correlated noise, the Video8 was designed for low $1/f$ noise and low channel-to-channel crosstalk. Table 1 summarizes the performance characteristics of the Video8 architecture.

The Video8 incorporates 8 fully differential readout channels, each composed of a preamplifier and a pair of integrators (Figure 1). The 8 channels are grouped into two banks of 4, with each bank being serviced by a single off-chip 20 bit ADC. The integrators box-car filter the signal for optimum noise performance, then perform a sample and hold. The two integrators associated with a given preamplifier operate in ping-pong fashion, with one performing an integration while the other holds a prior integration result for multiplexed output to one of the two off-chip ADCs. Multiplexing four channels to each ADC results in ADC operation at 400 kHz. The amplifiers and other internal circuitry of Video8 operate with a 5V supply voltage, while a separate 3.5V supply voltage determines the digital input signal voltage range.

Analog switches at the inputs to the preamplifiers provide flexibility for the input of reference and test levels. The inputs to the preamplifier are capacitively coupled and the feedback is purely capacitive, with a reset switch in parallel. Closing the reset switch nominally brings the preamplifier outputs to 2.5V. Additional capacitor/switch networks at the preamplifier inputs allow the preamplifier output levels to be adjusted, post-reset, in preparation for operation. The outputs of each preamplifier are connected through off-chip resistors to the inputs of a pair of integrators (see Figure 1). The feedback for the integrators is again purely capacitive with an associated reset switch in parallel. Switches at the integrator outputs multiplex the integrator signals for off-chip analog-to-digital conversion. The Video8 switches may be sequenced by external logic for optimal performance. Using off-chip resistors allows for some adjustment of the overall channel gain. With the use of low temperature coefficient (TC) resistors, the feedback capacitors ($\sim$ 20ppm/°C) dominate the temperature susceptibility.

The Video8 devices are manufactured using the C5N 0.5 $\mu$m CMOS process at Onsemi (ON Semiconductor), the same process used for manufacturing several ASICs designed at Caltech for prior space missions. Guard rings around groups of NMOS and PMOS FETs provide a "rad-hard by design" feature that boosts the latch-up threshold to >80 MeV/(mg-cm2), verified in test with heavy ions. The total dose tolerance has been verified up to 20 krad, well beyond the 7.5 krad SPHEREx lifetime requirement. The Video8 die size is 3.3x3.3 mm, and each device is packaged in a standard 84 pin ceramic quad flat pack. Caltech and the NASA Jet Propulsion Laboratory worked closely on the Video8 space qualification, following the process for prior Caltech-designed space ASICs.

**Table 1.** Video8 Performance Characteristics

| Parameter | Value |
|---|---|
| Amplifier bandwidth | 20 MHz |
| Read noise | 1-2 e |
| Gain stability | 20 ppm |
| Offset stability | 3 $\mu$V/C |
| Integral non-linearity | <0.07% |
| Channel-to-channel cross-talk | <2e-4 |
| Room temperature preamp output drift rate | <1 mV/s |

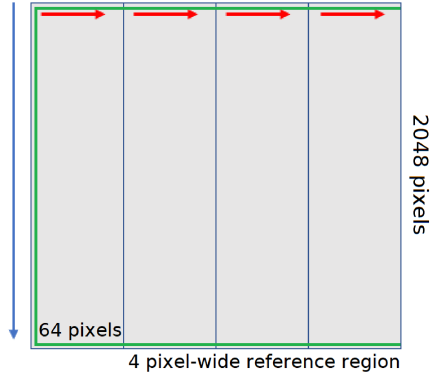

**Figure 2.** Illustration of the H2RG readout operated in 32-channel mode, truncated to 4 channels. Each channel is read simultaneously, row-by-row, progressing downward. The array has a surrounding border of reference pixels 4 pixels wide. We vary the sequence of row order for improved noise stability.

### 2.3. *Row-chopping*

Intensity mapping measurements of the extragalactic background require stable noise performance on tens of arcminute spatial scales where linear clustering peaks. However, noise stability is less important on small scales due to prominent Poissonian fluctuations from undetected galaxies. A new method for reading out channel rows, "row-chopping", transfers correlated spatial noise from low to high spatial frequencies. Because noise at small spatial scales is largely subdominant to photon noise, row-chopping improves stability on large spatial scales.

Row-chopping reads out the rows of an image in a pre-defined non-sequential order. Rows can then be re-ordered after an exposure has been taken to recover the sky image. The result transfers $1/f$ noise power produced by the readout electronics to smaller spatial scales. The row-chopping algorithm we employ incorporates an integer skip parameter $S$ which controls the scale of the noise reduction provided (Figure 3):

1. Read the first row in the image.

2. Skip over $S-1$ rows without reading them. If this results in skipping over rows that are beyond the last physical row in the image, wrap around to the beginning of the image and continue counting, counting one additional time each time you wrap back to the first row.

3. Read the current row.

4. Repeat steps 1-3 with the same $S$ until every physical row on the array has been read once.

The SPHEREx readout electronics execute row-chopping on each of the 32 channels simultaneously during readout. The parameter $S$ controls the spatial scale to which noise is transferred, allowing for the selection of an $S$ that minimizes noise power at low spatial frequencies (see Section 4.2). Row-chopping effectively reduces spatial $1/f$ noise along the vertical direction of the array, perpendicular to the rows. $1/f$ noise along each row, unaffected by row-chopping, is already small due to the short time required to read each row.

### 2.4. *Reference Pixels*

The largest remaining large-scale effect after row-chopping is an overall offset present within each channel due to common $1/f$ noise in the channel readout. We can use reference pixels to remove these per-channel offsets. Each H2RG array contains a 4 pixel-wide frame of the outermost pixels in the overall 2048×2048 grid which are not sensitive to light. These reference pixels are meant to reproduce the electrical behavior of an optical pixel without any photoresponse or photon noise. We use these pixels to correct for channel-level offsets from residual $1/f$ noise common to a channel readout.

These reference pixels are most easily read out in physical order, leaving 8 rows of reference pixels for each channel. There are an additional 4 columns of reference pixels on the right and left sides of the frame (only in channels 1 and 32).

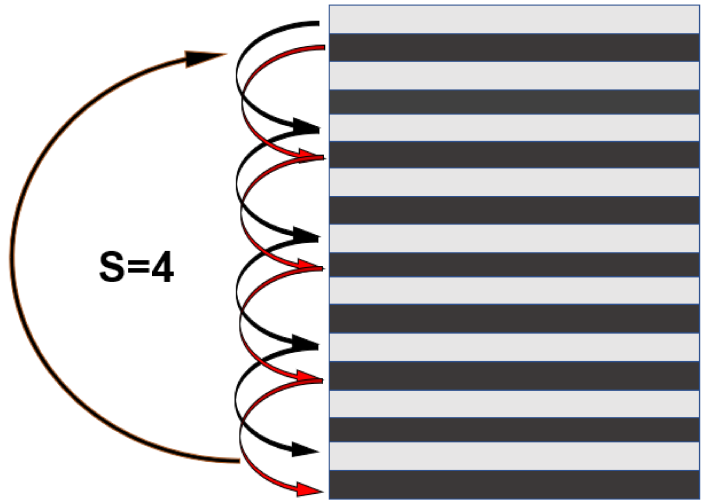

**Figure 3.** Illustration of row-chopping on an example grid for $S = 4$, with black arrows representing the first pass through the array and red arrows representing the second. A pre-defined number of rows (here, 4) are skipped over after reading each row, with a return to the first unread row every time the bottom of the array is reached.

To optimize offset correction, the SPHEREx readout system can sample reference rows additional times per image, interspersing them evenly through an exposure (see Section 4.3). These additional reads are intended to increase the accuracy of the reference pixels through statistical averaging.

In some cases we also note a separate offset between even and odd columns within a channel (see Section 3.2). While this even/odd offset only enters at high spatial frequencies and is therefore not relevant to measurements at large spatial frequencies, it can be corrected by using even/odd reference pixels.

### 2.5. Phantom Pixel Correction

Each Video8 periodically samples a DC reference from the bias supply, injected into the data stream as a "phantom pixel". The phantom pixel data are processed by the same sample-up-the-ramp algorithm as the H2RG pixels, correcting preamplifier $1/f$ noise, leakage current, and temperature drift. The Video8 leakage current is temperature-dependent and significant, typically a few electrons per second. Phantom pixel samples accurately remove these effects because only the low amplifier noise is present when sampling the reference.

During readout, we collect $c$ phantom measurements at the output of the Video8 with the input connected to a stable low-noise reference voltage. These phantom measurements are then averaged together into a single phantom pixel per row. Following this, we read the whole row of 64 pixels. We correct rows of the output image by subtracting the subsequent phantom pixel from each pixel in a row, resulting in 32×2048 corrections, as each H2RG channel has its own set of independent phantom pixels.

### 3. NOISE SIMULATION

#### 3.1. Simulation Inputs

In order to assess various correction parameters and to provide simulated noise images for the SPHEREx sky simulator package, we developed a Python simulation to generate unique noise images from predefined noise sources. We first characterize separately the noise power spectra produced by the various components of the readout electronics. Once power spectra are defined, we can generate an arbitrary quantity of simulated noise timestreams for a simulated image array. We generate an array of frames in this way for a simulated exposure and fit slopes for each pixel to produce a sample-up-the-ramp image.

We found that the SPHEREx readout noise timestream can be modeled by a noise power spectrum function $p$ of the form:

$$p = \alpha + \beta/f + \gamma/f^{1.3}. \tag{1}$$

The factors $\alpha$ and $\beta$ correspond to the white and $1/f$ noise coefficients of the system, respectively. Here, $\gamma$ is only used to model a component of excess noise in the readout integrated circuit (ROIC) amplifier, with the additional power of 1.3 being found to best match the measured noise data.

### 3.2. *Generating Simulated Images*

We determined the noise coefficients for each of the readout components by fitting $\alpha$, $\beta$, and $\gamma$ to a measured noise power spectrum (Table 2). Power spectra corresponding to the different noise components are show in Figure 4. We measured the ROIC amplifier white noise by correlated double sampling due to this noise component overwhelming other sources of noise at high sampling rates. For the Video8 and ROIC noise sources, we generate independent unique noise timestreams for each of the 32 H2RG channels.

Since the noise components of the bias supply act simultaneously across all channels, we generate a single bias timestream for the entire 32-channel image at the start of each simulation. We further split bias contributions into three components to model how they affect phantom pixel measurements (see Section 2.5). These consist of a "phantom uncorrelated" bias term which is only visible to the phantom pixel reads, a "phantom uncorrelated" bias term which is only visible to the array and not to the phantom pixels, and a "phantom correlated" bias term that is visible to both. There terms account for the fact that the phantom reference and the detector bias are partially correlated, with independent and common-mode noise. The magnitude of the independent components largely determines how effectively the phantom pixel correction method in Section 2.5 corrects for bias fluctuations in addition to its primary function of removing Video8 amplifier leakage current.

**Table 2.** Measured best-fit noise components for SPHEREx readout electronics

| Noise Source | $\alpha$ | $\beta$ | $\gamma$ |
| --- | --- | --- | --- |
|  | e$^2$/Hz | e$^2$ | e$^2$ * Hz$^{0.3}$ |
| Video8 | 3.71e-5 | 8.40e-3 | 0 |
| ROIC amplifiers | 1.97e-3 | 0 | 1e-1 |
| Phantom correlated bias | 0 | 8.80e-5 | 0 |
| Phantom uncorrelated bias, visible to phantom pixels | 0 | 3.64e-6 | 0 |
| Phantom uncorrelated bias, invisible to phantom pixels | 2.91e-6 | 3.07e-7 | 0 |

Note—Coefficients are calculated assuming a one-sided power spectrum convention

With noise coefficients for all sources defined, we produce unique timestream realizations. Let $R$ represent a random number with a uniform distribution on $(0, 1)$ and let $f_s$ be the sampling frequency of the system (100 kHz). Given an arbitrary $n \times 1$ two-sided power spectrum array $\vec{p}$, $n$ odd, we produce a timestream realization $\vec{x}$ corresponding to $\vec{p}$ by applying random phases to the non-DC components of the two-sided power spectrum components and taking an inverse Fourier transform as follows:

$$\vec{A} = [p_0^{1/2}, p_1^{1/2}, ..., p_n^{1/2}], \tag{2}$$

$$\vec{\phi} = [e^{2\pi i R_0}, e^{2\pi i R_1}, ..., e^{2\pi i R_{\text{floor}(n/2)}}], \tag{3}$$

$$\vec{a} = \vec{A} \cdot [1, \vec{\phi}, \vec{\phi}^*], \text{ and} \tag{4}$$

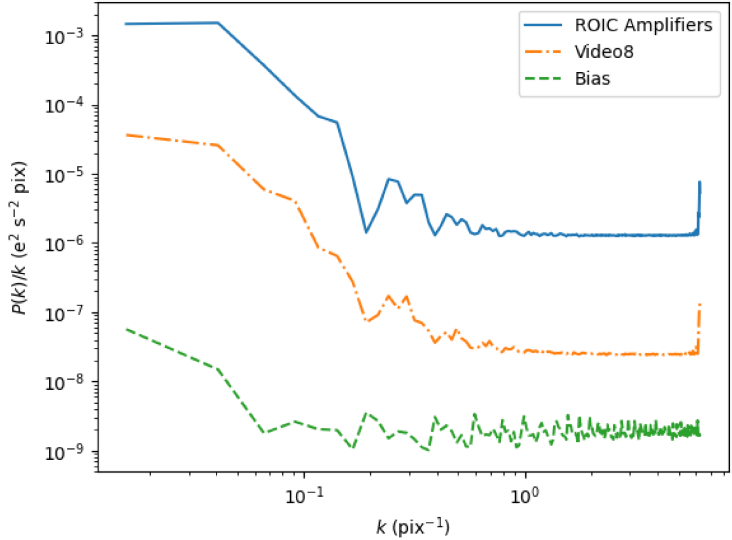

**Figure 4.** Simulated spatial power spectra realizations corresponding to each noise component power spectrum applied to a 2048×2048 image obtained in readout order, with no dark current. The ROIC amplifier noise dominates at all spatial frequencies. The prominent peak at high $k$ for the Video8 and ROIC spectra is caused by alternating column noise, and is not relevant for large-scale measurements.

$$\vec{x} = \Re(\mathcal{F}^{-1}[\sqrt{nf_s} * \vec{a}]). \tag{5}$$

Since this calculation depends on floor($n/2$) random numbers, we can generate a large number of relevant timestreams given enough computation time. Once we generate a unique timestream for each channel and combine it with the bias timestream component, we can produce a unique frame corresponding to the noise power present on a perfectly dark exposure with no fixed dark current. For a sequence of $N$ frames, we fit a slope to each pixel to produce a full noise image realization.

An additional source of H2RG noise is present in the form of an offset between the even- and odd-numbered columns of each channel, referred to by Rauscher (2015) as alternating column noise (ACN). ACN effectively results in two common offsets for each channel. ACN primarily produces spatial noise at scales significantly smaller than those relevant to large-scale background measurements, but we include it for completeness. We model ACN by splitting all noise components (except for the common bias components) into two realizations and applying one to each column parity, producing two independent offsets per channel.

### 3.3. *Unmodeled Noise*

Our measurements (see Section 5) also indicate the presence of a low level of low-frequency per-pixel telegraph noise. Such a noise component increases the effective white noise level in long integrations. This same noise also reduces the efficiency of multiple reference sampling once the readout noise falls below the per-pixel noise. We have chosen not to model this per-pixel noise, but point out its effects in long integrations, where it causes an offset in noise levels between simulated and measured images.

### 4. NOISE REDUCTION AND OPTIMIZATION

After producing simulated noise images as detailed in Section 3, we apply the following noise reduction techniques on simulated images, calculate their spatial power spectra, and compare the post-processed power spectra with the spectra of the original images.

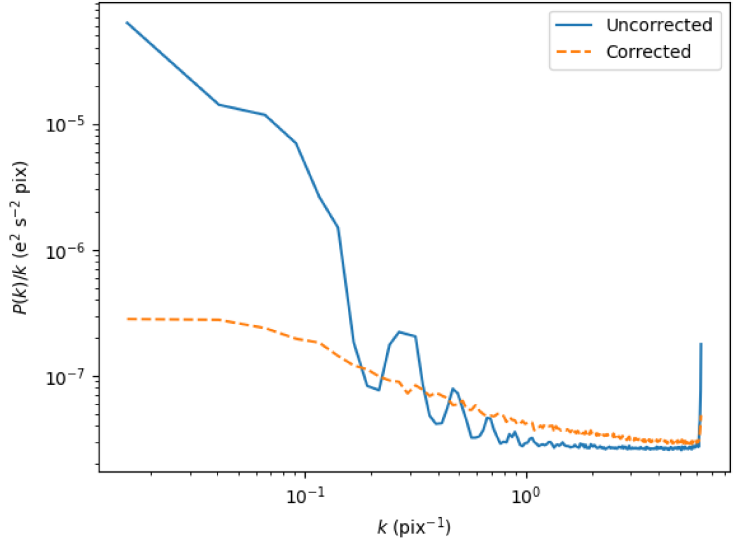

**Figure 5.** Power spectra for a 32-channel simulated image with all noise sources active except the ROIC amplifier terms, before and after $c = 4$ phantom pixel correction. The $1/f$ component here is from Video8 and bias noise. While subdominant to the ROIC noise, these sources of $1/f$ noise are mitigated by phantom pixel corrections.

### 4.1. Phantom Pixel Correction Performance

For the particular noise sources monitored by the phantom pixel reads, phantom pixel corrections are very effective at reducing spatial $1/f$ noise due to the low Video8 noise. For simulated images generated using only noise sources visible to the phantom pixels and using $c = 4$ co-adds, the spatial noise is significantly reduced at low spatial frequencies (see Figure 5). For a full SPHEREx readout simulation with all noise sources active, the noise power contributed by the Video8 and bias is subdominant to the ROIC noise. However, this correction remains critical to remove the large leakage current from the Video8 amplifiers.

### 4.2. Row-chopping Optimization

The spatial scale for noise reduction in the row-chopping operation is determined by the parameter $S$, the number of rows skipped during each row read. This value is selected during system development and is not changed during flight. For $S$ much less than the total number of rows in the image (in this case, 2048), increasing $S$ generally pushes noise to smaller and smaller spatial scales.

As $S$ increases, some noise wraps around to larger spatial scales rather than continuing the initial downward trend. Thus there is an optimal $S$ that depends on project requirements. In this paper, we selected a skip parameter of $S = 24$ to minimize noise at tens of arcminutes scales (see Figure 6 and Figure 7). We selected this value by applying row-chopping to a full simulated image with the parameters described in Table 2. Increasing $S$ beyond a few dozen rows results in a slight increase in noise at large spatial scales due to noise wrapping.

### 4.3. Reference Correction

Row-chopping mitigates $1/f$ noise in the vertical direction across the array but not across channels due to current offsets in individual H2RG channels, which are created by independent $1/f$ noise in each channel over an integration. Additional current offsets between even/odd columns at smaller spatial scales are produced by ACN. Both of these offsets can be eliminated by filtering the horizontal axis of the image centered in Fourier space, but at the cost of losing some astrophysical information. We can use reference pixels to approximate these corrections without this downside.

The most basic method for reference correction is to subtract a single average of all of the reference pixels in each channel. Because the H2RG array uses a 4-pixel-wide frame of reference pixels, this results in $8 \times 64 = 512$ reference pixels available for each channel, with an additional 8176 reference pixels for the first and last channels distributed along the side of the channel. More complex methods for reference correction have also been developed (see Rauscher et al. (2017) and Kubik et al. (2014)).

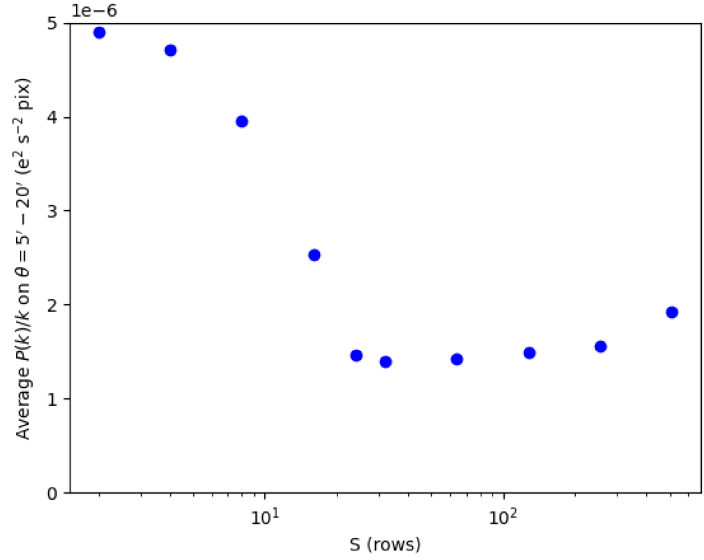

**Figure 6.** Average noise power on $\theta = 5 - 20$ arcminute scales ($0.03 < k < 0.13$ pix$^{-1}$) with varying skip parameter $S$ for a 32-channel $N = 74$ simulated reference-corrected image with all noise sources active. Here, we calculate noise after excluding the horizontal axis in Fourier space (see Figure 7) to isolate the effect of row-chopping from reference correction limitations.

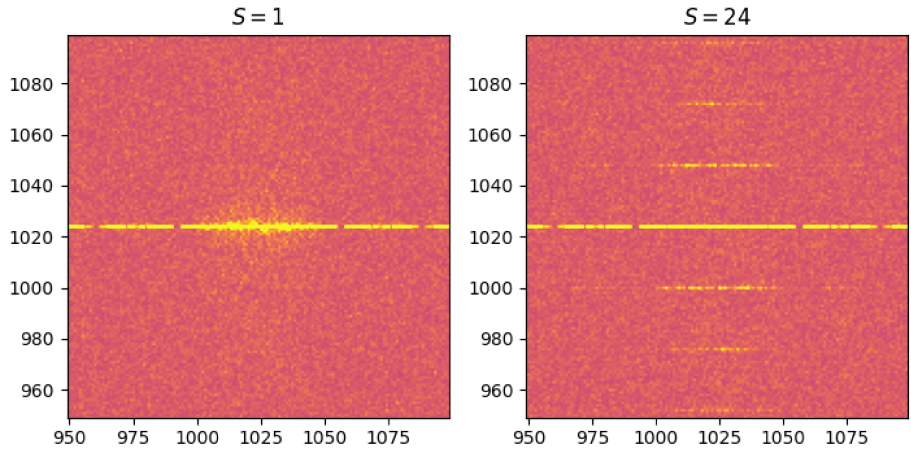

**Figure 7.** 2D FFT magnitudes in arbitrary units of a 32-channel simulated image before and after $S = 24$ row-chopping, cropped to show the region near the origin. Noise power is transferred from the horizontal axis to discrete packets above and below. Significant power remains on the horizontal axis, demonstrating the need for reference correction. We excluded the horizontal axis in Figure 6 to optimize the choice of skip parameter.

We improved on the default reference correction method with two strategies: (1) reading the top and bottom rows repeatedly in place, and (2) spreading the multiple reads throughout the image. As shown in Figure 8, distributing the reference reads improves the performance of the correction. This improvement is due to low-frequency components in the noise timestreams, which are mitigated by reducing the average time interval between reference samples.

The overall reference correction algorithm operates as follows, utilizing only the reference rows physically at the top and bottom of each channel:

1. Define a desired number of reference row reads per channel $r$.

2. Define positions for the reference rows such that they are evenly distributed throughout the channel and insert these rows into the readout order by re-reading a physical reference row at each position. Physical reference rows are read cyclically, resulting in each being read an average of $r/8$ times per frame.

3. After producing a full image, average together all reference pixels for each channel, in even and odd columns (resulting in two corrections per channel) if desired, and subtract those values from the respective channel. A $\sigma$-clip operation should be performed prior to averaging reference pixels to eliminate any outlier hot pixels.

By default, we subtract a single average from each channel and do not correct for ACN in order to increase the statistical weight of the large spatial scale correction. Since the lowest frequency variations in the channel timestreams are effectively eliminated by $S = 24$ row-chopping, subtracting a single average from each channel is essentially identical to subtracting a best-fit average to the reference pixels.

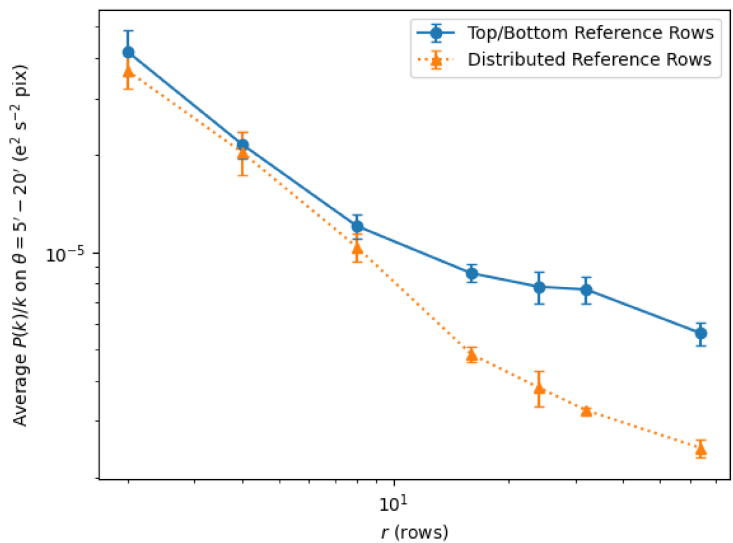

**Figure 8.** Average noise power on $\theta = 5 - 20$ arcminute scales as a function of the number of row samples $r$ for both reference correction methods with $S = 24$ row-chopping, averaged over 5 simulated 32-channel images. The distributed method provides an improvement over repeated sampling of the default reference rows, at little cost to readout time from row shifts.

**Table 3.** Simulation Input

| Parameter | Value |
|---|---|
| Number of Frames $N$ | 74 |
| Time Per Frame $T_{int}$ (s) | 1.51 |
| $S$ (rows) | 24 |
| $r$ (rows) | 32 |
| $c$ (pixels) | 4 |

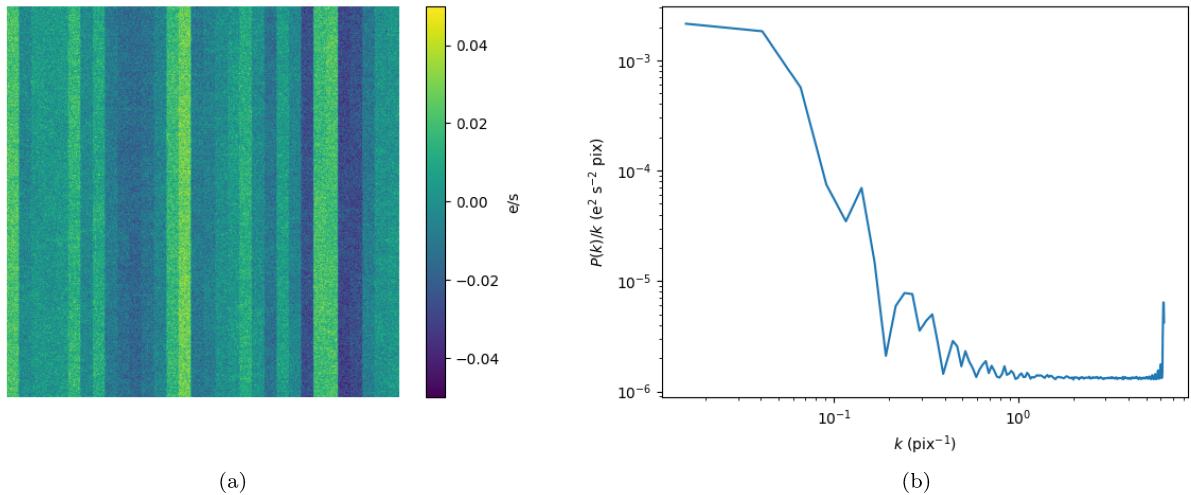

(a)             (b)

**Figure 9.** (a) Example output 32-channel $N = 74$ noise realization generated by the described simulation and (b) corresponding spatial power spectrum, with phantom pixel correction but without reference pixel correction. Large channel offsets due to low-frequency $1/f$ noise are visible prior to reference correction.

## 5. MODEL PREDICTIONS COMPARED WITH MEASUREMENT

### 5.1. *Simulation Output with Optimized Parameters*

The described simulation generates noise image realizations that approximate the noise components present in dark images. Table 3 shows the simulation inputs for a standard SPHEREx exposure. Figure 9 shows an example of an image produced by the simulation and its corresponding spatial power spectrum.

### 5.2. *Row-chopping and Reference Correction Model Predictions*

While phantom pixel correction is largely negligible on simulated images that include all noise sources due to the already extremely low Video8 noise components, row-chopping and reference correction applied together have a very significant effect on correlated spatial noise. Using the selected $r = 32$ reference correction alongside $S = 24$ row-chopping, the average spatial noise on $\theta = 5 - 20$ arcminute scales is brought well below the original level on simulated image realizations (see Figure 10).

### 5.3. *End-to-end Measurements*

We compared simulated images with dark images produced by a representative H2RG array and readout electronics. Measured dark images were collected at an array temperature of 40K. Row-chopping was implemented at $S = 24$ rows during readout. The detector has a small level of optical response during these tests due to a combination of multiplexer glow and residual light leaks. We used phantom correction at $c = 4$ to remove Video8 leakage and amplifier $1/f$ noise.

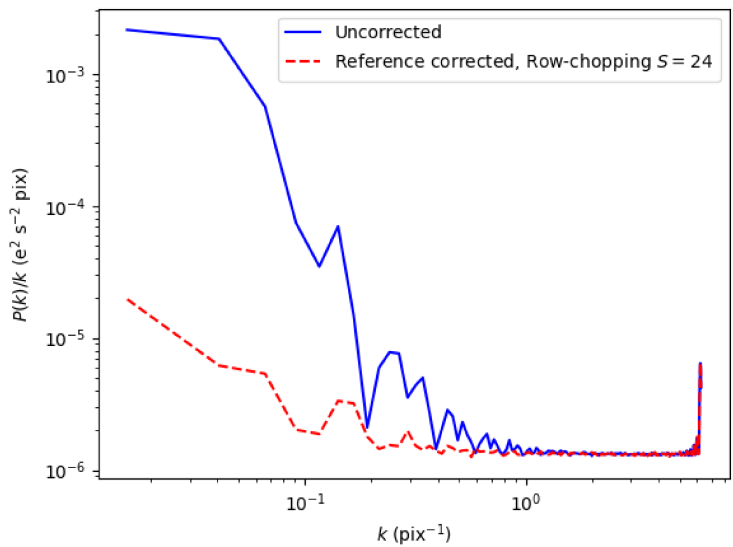

**Figure 10.** Spatial power spectrum from a 32-channel $N = 74$ simulated image realization with $S = 24$ row-chopping and $r = 32$ reference correction. We obtain a significant improvement at low $k$ compared with the uncorrected image.

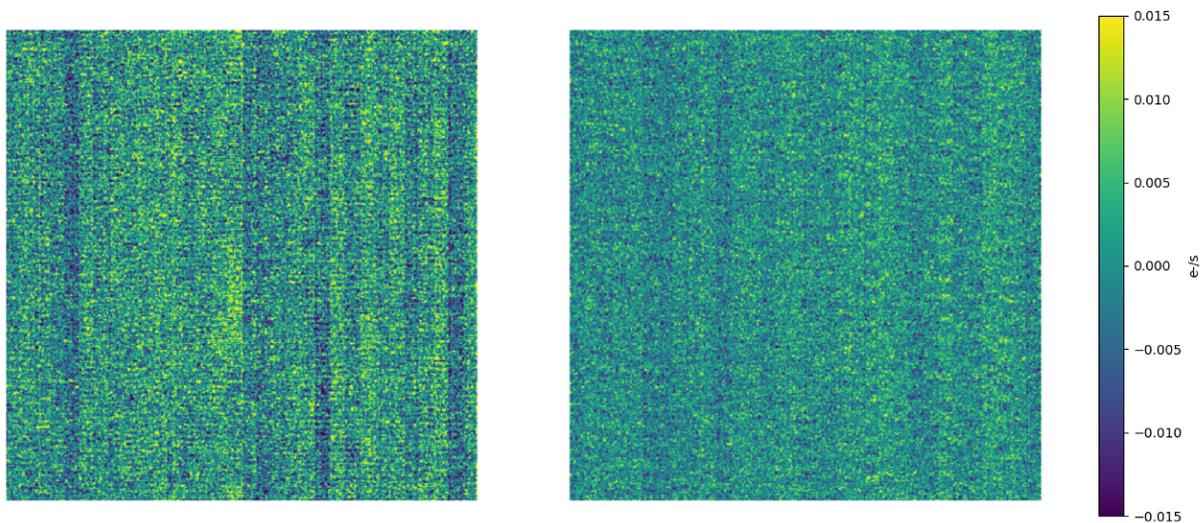

**Figure 11.** Side-by-side comparison of a measured (left) and simulated (right) $N = 74$ 32-channel image pair difference. Each image has $S = 24$ row-chopping and $r = 32$ distributed reference correction.

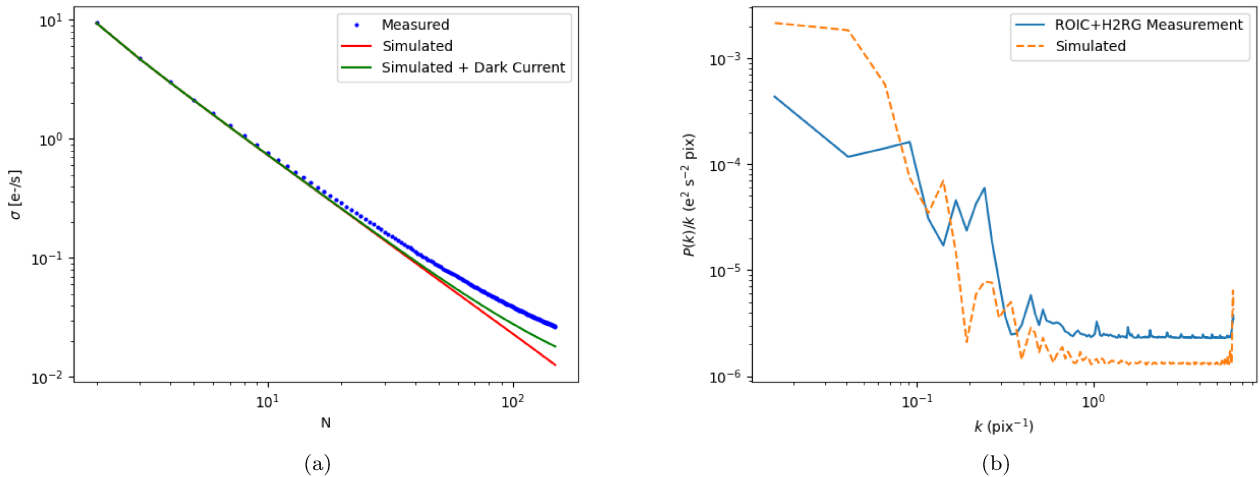

(a) (b)

**Figure 12.** (a) Pixel standard deviation $\sigma$ as a function of frame count $N$, after $\sigma$-clipping, for measured and simulated images produced using a single Video8 (8 channels) with $r = 32$ ACN reference correction and (b) spatial power spectra for uncorrected 32-channel image pairs at $N = 74$, showing an offset between output spectra, particularly at high $k$. For (a), measured channels are processed by subtracting a low-noise $N = 386$ ACN reference-corrected long exposure. Unmodeled per-pixel noise increases the deviation between measured and simulated channels at large $N$, even after accounting for dark current variation (see Section 3.3).

We compare images at $N = 74$ frames, the frame count planned for flight. Measured images were processed prior to reference correction by subtracting a second $N = 74$ exposure to remove small constant dark current offsets in the testbed, followed by iterative $\sigma$-clipping to 5 standard deviations to eliminate hot pixels and an additional factor of $1/\sqrt{2}$ to account for the image difference. A side-by-side comparison of a measured and simulated image is shown in Figure 11. The simulated images model the measured images well at low $N$ before deviating as $N$ increases. At $N = 74$, the apparent white noise level of the measured images is approximately 30% higher than in the simulated image after dark current is added (see Figure 12). We attribute this departure to per-pixel $1/f$ telegraph noise (see Section 3.3). The behavior at low $k$ is otherwise well modeled by the simulation before applying reference correction.

### 5.4. System Performance at Large Spatial Scales

Implementing $S = 24$ on-board row-chopping and $r = 32$ reference correction with distributed reference row reads on the physical array and readout electronics results in a significant reduction of noise power on large spatial scales. For the specific application to SPHEREx, a relevant comparison can be made with the photon noise level caused by the bright Zodiacal light foreground, which contributes a Poissonian noise spectrum. While the SPHEREx instrument noise power requirement is not based on photon noise, the level of the ZL photon noise spectrum acts as a lower limit on possible instrument observed noise power.

Figure 13 demonstrates the reduction in spatial power on $\theta = 5 - 20$ arcminute scales resulting from the described noise reduction methods. The noise reduction on measured images is significantly less than the reduction seen in simulated images (see Figures 10 and 13). Our testbed dark image measurements at various row-chopping $S$ values also confirm the optimal $S$ of a few dozen rows indicated by the simulation (see Figure 14). Measured images also show that increasing the number of reference row visits does not completely eliminate the excess channel offsets, with less noise reduction for higher $r$ than simulated images would suggest. We note that the reference pixels do not completely follow the behavior of the optical pixels, even in the absence of photon noise in testbed images. Figure 13 includes an example of an alternative post-processing method for reference correction subtracting the median of all pixels (optical and reference) in each channel instead of the average of reference pixels, which we briefly mention here for the sake of comparison. Median correction in this way results in improved noise performance, with resulting spectra dropping below the SPHEREx band 1-4 photon noise levels at all presented spatial scales. However, this improved performance over the use of built-in reference pixel correction comes at the cost of loss of astrophysical information on real sky images, as each channel will necessarily include sky features. We further explore the use of median correction and the efficacy of SPHEREx reference pixel offset correction, as well as provide a more detailed overview of the testbed, in a future publication.

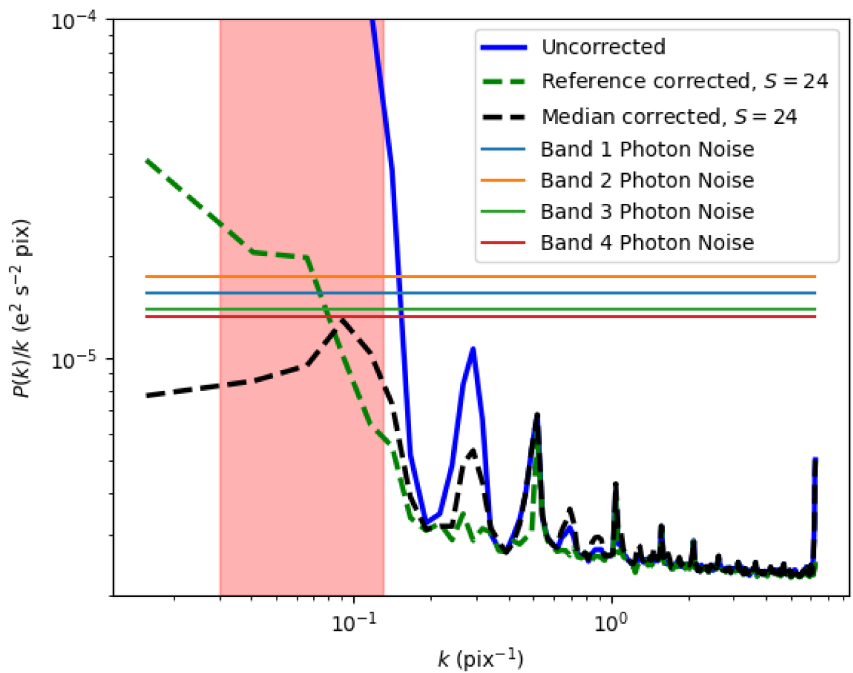

**Figure 13.** Spatial power spectrum of a 32-channel $N = 74$ image pair difference obtained with a physical H2RG array and the readout electronics, before and after $r = 32$ reference correction with distributed rows. $S = 24$ row-chopping is applied during readout. The range of $\theta = 5-20$ arcminutes is shown in red, corresponding to $0.03 < k < 0.13$ pix$^{-1}$. An estimate of the photon noise spectra due to the expected Zodiacal light incident photocurrents in the first 4 SPHEREx bands **(spanning 0.75-3.8$\mu$m) for the SPHEREx deep fields at the ecliptic poles is also included**. Reference correction and row-chopping provide a significant noise reduction at large spatial scales on measured images, with spatial noise brought near the level of the ZL photocurrent photon noise.

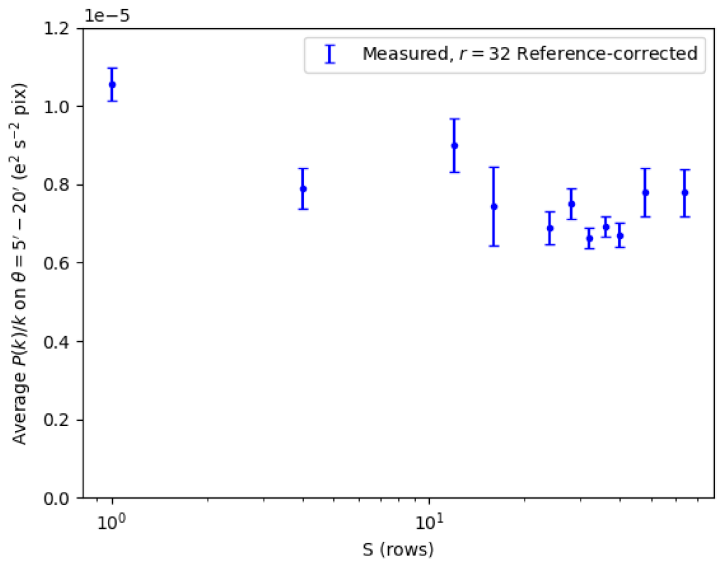

**Figure 14.** Average noise power on $\theta = 5-20$ arcminute scales ($0.03 < k < 0.13$ pix$^{-1}$) with varying skip parameter $S$ for 5 measured 32-channel $N = 74$ image pair differences with $r = 32$ reference correction. The measured optimal $S$ of a few dozen rows is consistent with that provided by the simulated noise realizations, though with significantly more variation and a higher overall noise level. A substantial decrease in large-scale noise power is observed even at small values of $S$.

## 6. CONCLUSIONS

We developed new algorithms for improving large-scale noise properties in H2RG detector arrays for intensity mapping applications. A combination of reference correction and row-chopping substantially reduces the noise power present at large spatial scales. Row-chopping provides excellent low-frequency noise stability along the vertical direction of the array without appreciably increasing readout time. The custom Video8 preamplifier developed for array readout produces extremely low $1/f$ noise power using a phantom pixel correction scheme. As these methods are implemented during array readout rather than post-processing of sky exposures, no significant additional wavelength-dependence or effect on point sources is expected. While this work focuses specifically on applications for SPHEREx and H2RG arrays, the described row-chopping technique could be applied to any row-by-row infrared imaging system where reduced noise power at large spatial scales in desired.

The methods we developed reduce spatial readout noise to near the level of photon noise due to Zodiacal light photocurrent on large spatial scales for SPHEREx, with a $\sim 1$ e/s photocurrent and a 112s integration time. The combined large-scale statistical noise seen by the instrument resulting from the described noise-reduction methods is expected to be well below predicted EOR signals (Doré et al. (2014), especially Figure 24). Noise stability at large scales is limited by per-pixel telegraph noise in the H2RG built-in reference pixels, which limits the effectiveness of repeatedly sampling reference pixels. Our simulated noise realizations generated from constituent noise components closely approximate measured images for low frame count $N$, but deviate at high $N$ (particularly at small spatial scales). We plan to optimize the readout parameters in a future work using measured array data.

## 7. ACKNOWLEDGEMENTS

A portion of the research described here was conducted at the Jet Propulsion Laboratory (JPL), California Institute of Technology. This article and analysis made use of the Astropy (Astropy Collaboration et al. (2022)), Numpy (Harris et al. (2020)), and Matplotlib (Hunter (2007)) Python packages. We acknowledge support from the SPHEREx project under a contract from the NASA/GODDARD Space Flight Center to the California Institute of Technology.